\documentstyle[preprint,aps]{revtex}

\begin{document}
\draft
\title{From surface to random criticality in layered planar Ising models.}
\author{Lev V. Mikheev}
\address{Nordita, Blegdamsvej 17, DK-2100, Copenhagen \O, Denmark}
\date{March 1994}
\maketitle
\begin{abstract}
A general case of a spatially nonuniform planar layered Ising model, or an
equivalent quantum Ising chain, is analysed with an exact functional real space
renormalization group. Various surface, finite size, quasiperiodic and random
layer (McCoy-Wu) universality classes are obtained and discussed within a
single theoretical framework leading to new insights into the nature of random
criticality.
\end{abstract}

\narrowtext

While equilibrium critical phenomena in spatially uniform systems are gradually
becoming a standard textbook chapter, much less is known about the way the
critical state of matter is altered by the presence of spatial inhomogeneities
of a general form. Models with uniformly distributed point {\em disorder}, on
one hand, and with a planar {\em surface}, on the other, have attracted
considerable attention and considerable progress in understanding of both has
been achieved. However, the developments along these two directions have been
largely independent, leaving a whole range of inhomogeneities largely
unexplored. Important examples falling into this range are, to name just two,
disordered quantum systems \cite{dsf}, in the path integral representation of
which the disorder is fully correlated in the time direction, and fluids in
porous media \cite{Moses_Chan}, in which case disorder enters in the form of
random {\em surfaces}.

Here I am attempting to bridge this gap by presenting a new analysis of the
class of layered planar Ising (LPI) models. A layered Ising model
\cite{levfish2} is defined as having its bond strengths depending in an
otherwise arbitrary way on just {\em one} of the $d$ spatial coordinates, $z$;
in each layer perpendicular to that coordinate the bonds are of the same
strength. These models may be viewed as a natural step in generalization of
models with one planar surface towards the complexity of real systems. In
particular, any LPI model is equivalent to an $s=1/2$-quantum Ising chain in
transverse field which, depending on the way the transfer matrix is applied,
can be made inhomogeneous either in space or in time. Besides, the rapidly
developing technology of artificially grown multilayers provides natural
experimental realizations of the layered geometry. Starting from the pioneering
work of McCoy and Wu \cite{McCoyWu}, a number of LPI models, both with surfaces
and randomness \cite{NO}, have been solved exactly. Subsequent scaling analysis
of the exact solutions for surface and film geometries has played a crucial
role in formulating surface \cite{BH,VLB} and finite size scaling \cite{FP}.
More recently important progress has been made towards understanding of the
random layer model \cite{SM,dsf}, so the whole class seems to be a good testing
ground for unifying ideas.

This work builds upon the {\em microcanonical} density functional method,
proposed recently by M. E. Fisher and the author. Exact Euler-Lagrange
equations obtained for the LPI class in \cite{levfish2} are used to construct
a functional renormalization group (RG) flow in the space of energy density
profiles. Solving for the fixed points of this flow generates a list of
universality classes found in LPI models. Common features of the flow in the
vicinity of the fixed points describing the free film and random criticalities
leads to interpretation of the latter as a hierarchy of thin-thick film
crossovers with a significant element of quasi-one-dimensional behavior. This
qualitative insight allows then for quantitative predictions to be made
regarding critical behavior in quasiperiodic and long-range-correlated random
sequences.

The Euler-Lagrange equations found in \cite{levfish2},
\begin{equation}
q\frac{\varepsilon_q(z)-(2q)^{-2}\ddot{\varepsilon}_q(z)}
{[1-\varepsilon_q^2(z)+(2q)^{-2}\dot{\varepsilon}_q(z)^2]^{1/2}}=t(z),
\end{equation}
describe the response of the energy density components $\varepsilon_q(z)$ to
the scaled temperature field $t(z)$, which characterizes deviation from
criticality of the local bond strengths in a given LPI model. Here $q$ is the
wave number in the direction parallel to the layers, $\varepsilon_q(z)$ is the
contribution to the total energy density from the excitations belonging to the
$q$-sector of the equivalent fermion problem \cite{SM,levfish2};
$\dot{\varepsilon}_q(z)\equiv d\varepsilon_q/dz$, etc.  An inverse latttice
constant has been absorbed into the definition of $t$, giving it the proper
scaling dimension of inverse length; the normalization of
\cite{levfish2,onsager} makes the absolute value $|t|=\xi^{-1}(z)$, where
$\xi(z)$ is the correlation length of a uniform model with the same bond
strengths as the given $z$-layer (recall that $\xi\propto |t|^{-1}$ in a
uniform planar Ising model).  In this formalism the vertical bonds (connecting
subsequent layers) have to be ferromagnetic, but the horizontal bonds can take
either sign thus allowing for restricted forms of frustration (cf. \cite{SM}).
Eqs. 1 are exact in the scaling limit $t(z)\ll\Lambda$, with the cutoff
$\Lambda$ being of the order of the inverse lattice spacing. They have to be
solved for every $q$ between $0$ and $\Lambda$; after that the energy density,
free energy, etc., are obtained by integration over $q$ of simple functions of
$\varepsilon_q(z)$ and $\dot{\varepsilon}_q(z)$. This program has been
explicitly carried out for simple surface \cite{levfish2} and superlattice
\cite{onsager} geometries, as represented by step-function and periodic $t(z)$,
respectively.

In the latter case, apart from the information obtained from the explicit
expressions for the free energy and total energy density profiles, it turned
out to be instructive to study evolution of the {\em partial} energy density
profiles $\varepsilon_q(z)$ in the long wave length limit $q\rightarrow 0$;
that evolution resembled many fetures of a functional renormalization group.
Indeed, taking a partial derivative of the left hand side of (1) with respect
to the logarithmic length scale along the layers, $l=\ln(\Lambda/q)$, one
obtains a partial differential equation depending only on $\varepsilon_q$, its
first two spatial derivatives, and, generally, on $q=\Lambda e^{-l}$ as well.
In this approach, the field $t$ can be viewed as fixing the initial data at
large $q=\Lambda$, where $\varepsilon(z;l=0)\approx t(z)/\Lambda$. The
evolution of profiles as the logarithmic length scale $l$ increases follows
then its own dynamics, independent of the source $t$, which, according to (1),
becomes an integral of motion for this flow. The simplest illustration is
provided by the uniform limit $t=\mbox{const},
\partial\varepsilon/\partial z=0$. Differentiating both sides of (1) with
respect to $l$ gives then
\begin{equation}
\partial\varepsilon/\partial l=\varepsilon(1-\varepsilon^2).
\end{equation}
As anticipated, Eq. 2 has three fixed points: $\varepsilon=0, \pm1$. In the
scaling regime, $t\ll\Lambda$, the flow always starts in the vicinity of the
ultraviolet-stable critical fixed point $\varepsilon=0$ and then takes the
system either to the high-temperature, $\varepsilon=1$, or to the
low-tmeperature, $\varepsilon=-1$, fixed point, depending on whether the
initial value $\varepsilon(l=0)=t/\Lambda$ is positive or negative,
correspondingly. The eigenvalue $\lambda_t=1$ governing the flow out of the
critical fixed point,
$\partial\varepsilon/\partial l\approx\lambda_t\varepsilon$, gives the correct
value of the correlation length exponent $\nu=1/\lambda_t=1$.

The general case of position dependent $t$ requires supplementing flow in $q$
with rescaling $\varepsilon\rightarrow\varepsilon\exp(-\omega_{\varepsilon}l),
\  z\rightarrow z\exp(-\omega_zl)$, where the scaling dimensions
$\omega_{\varepsilon},\ \omega_z$ are tuned to achieve nontrivial fixed point
structure and make the RG equations $l$-independent. In pracitice, it is
easier to work with the integral of motion (1): each RG fixed point profile
\begin{equation}
\varepsilon_q(z)=q^{\omega_{\varepsilon}}{\cal E}(q^{\omega_z}z)
\end{equation}
has to satisfy (1) with the same $t(z)$ for every small $q$. A straightforward
analysis yields several fixed points.

First, at $\omega_z=0, \omega_{\varepsilon}=1$ one finds the profile
\begin{equation}
{\cal E}(y)=-\int^{y}dz\sinh(g(z)),
\end{equation}
describing relaxation towards the uniform critical fixed point profile
$\varepsilon\equiv 0$. The {\em thermal potential}
\begin{equation}
g(z)=2\int^{z} t(z_1) dz_1,
\end{equation}
will play a crucial role below. The condition of criticality in a layered
Ising model discussed in \cite{levfish2} can be made more precise now: for (8)
to make sense there has to exist a choise of integration constant in (5) such
that the integral in (4) is bounded for all $y$ (note that $\varepsilon$
cannot exceed unity). This rather stringent constraint is satisfied in the
critical state of a periodic \cite{onsager} and, as one will see below, simple
quasiperiodic LPI models, but not at the ferromagnetic transition in the
random layer model, where $g(z)$ performs a random walk.

If positive values of $t$ dominate, so that at $z\rightarrow\pm\infty$ the
potential $g$ diverges as $\pm |z|$, respectively, then the profile relaxes
towards the high-temperature fixed point $\varepsilon=1$. Similar relaxation
to $\varepsilon=-1$ occurs for mostly negative $t$; the two can be combined
into the scaling form $\varepsilon=\pm 1\mp q^2{\cal E}_{\pm}(z)+o(q^2)$, where
\begin{equation}
{\cal E}_{\pm}=\frac{1}{2}\int^{z}_{-\infty}dz_1\int^{+\infty}_{z}dz_2\
e^{\mp[g(z_2)-g(z_1)]}.
\end{equation}
The scaling dimensions describing this {\em leading irrelevant} behavior at
the noncritical fixed points are evidently
$\omega_z=0,\ \omega_{\varepsilon}=2$.

If $t(z)$ is mostly positive at $z>0$ and mostly negative at $z<0$, so that $g$
has a global minimum around $z=0$, one finds a nontrivial fixed point
\begin{equation}
{\cal E}(y)=-1+2\int^{y}_{-\infty}e^{-g(z)}dz/
\int_{-\infty}^{+\infty}e^{-g(z)}dz
\end{equation}
scaling with  $\omega_z=\omega_{\varepsilon}=0$. It describes an interface
between the low-temperature phase on the left and the high-temperature phase
on the right; note that inversion of this solution, $z\rightarrow -z$, implies
inverting sign of $g$. In fact, by choosing finite instead of infinite limits
of the integrals in (6), (7) one still obtains a local solution to (1); those
local solutions will be used below in analyzing the case of a generic, sign
indefinite $t(z)$.

The remaining fixed point profiles describe response of a critical system to
surface/interface perturbations. The choise of $\omega_z=1/a>1,
\omega_{\varepsilon}=0$ yields a solution
\begin{equation}
{\cal E}(y)=(1+A^{-2}y^{2a})^{-1/2}
\end{equation}
describing {\em local}, $\varepsilon_q(z)=t(z)/(t^2(z)+q^2)^{1/2}$, response
to a power-law tail $t(z)=Az^{-a}$. Scaling in this case is somewhat spurious:
the local dependence of $\varepsilon_q$ on $t$ is realized whenever the source
$t$ varies slowly on the scale of the {\em local correlation length}
$\xi(z)=t^{-1}(z)$, i. e. $t^{-2}\dot{t}\ll 1$. This condition, bound to fail
near every zero of $t$,  yields $a<1$ for power-law sources $t\propto z^{-a}$.
Physically it implies that the essential correlations decay faster than the
bond strength changes, therefore any physical property can be calculated by
simply adding contributions from quasihomogeneous regions. The marginal value
$a=1$ leads one to the scaling dimensions $\omega_z=1,\ \omega_{\varepsilon}=0$
of the uniform critical model. Although an explicit solution of the fixed-point
equation
\begin{equation}
[{\cal E}-\frac{1}{4}\ddot{\cal E}][1-{\cal E}^2+\frac{1}{4}\dot{\cal
E}^2]^{-1/2}=q^{-1}t(y/q)
\end{equation}
for $q^{-1}t(y/q)=A/y$ is not available at the moment, presumably such solution
exists for any value of the dimensionless amplitude $A$; interesting properties
of this line of fixed points are discussed in \cite{VLB}. Besides
$t\propto z^{-1}$, the right hand side of (9) is $q$-invariant for $t(z)=0$ and
$t(z)=g_0\delta(z)$; the corresponding fixed point profiles are
\begin{equation}
{\cal E}_s=\mbox{const}\ e^{\pm 2y};\ \ {\cal E}_d=\tanh(g_0)e^{-2|y|}.
\end{equation}
The first one describes a surface of a critical half-plane in the absence of
long range perturbation in the bond strengths, including $t\propto z^{-a}$
with $a>1$ \cite{VLB}. Adding two of these half-plane solutions yields
${\cal E}_d$ describing response of a critical system to a defect line
\cite{levfish2}.

These fixed points allow for a compact descritption of a general case of an
interface geometry \cite{levfish2}. In the most generic case the interface
separates two noncritical half-spaces characterized by temperature fields
$t_1, t_2$. Only the relative sign of the $t_1$ and $t_2$ turns out to matter
then: $t_1$ and $t_2$ of the same sign inevitably lead to profiles flowing
into one of the uniform fixed points (6); opposite signs of the temperature
fields, on the other hand, lead to the flow into the nontrivial fixed point
(7). This discontinuous change in the form of the profile as one of the
half-planes passes its critical point is at the origin of the surface
{\em latent specific heat} found in the exact solution of the half-infinite
model \cite{McCoyWu,BH,levfish2}. At criticality of {\em one} of the
half-planes the flow can result in one of three fixed points, determined by
the {\em range} of the surface perturbation in the $t$-field, as explained
above and in agreement with the analysis in \cite{VLB}. Finally, when
{\em both} half-planes go critical simultaneously the total strength of the
bond perturbation in the interfacial reion, as encapsulated in the amplitude
$g_0=2\int tdz$ of the $\delta$-function, determines which of the line of
fixed points ${\cal E}_d$ (10) attracts the flow.

An insight into the nature of some of these fixed-point profiles is provided
by the exact equivalence \cite{SM,levfish2} between each $q$-sector of an LPI
model and a ficticious one-dimensional classical Ising chain at temperature
$T_{d=1}\sim|\ln q|^{-1}$ and nonuniform reduced magnetic field
$H_{d=1}(z)/k_BT_{d=1}\sim t(z)$. The energy density components
$\varepsilon_q$ map onto the magnetization density of the corresponding
$q$-chains. The low-$q$ limit of the planar model corresponds to the
low-temperature limit of the chain. The latter is known to be determined by
the statistics of a dilute gas of pairs of antiparallel spins which I will
call {\em kinks} below. A ``spin down-spin up'' kink, to which orientation
$\sigma=1$ will be assigned, has to be followed by an ``up-down'' one,
$\sigma=-1$, etc.  The energy of a kink at position $z$ along the $q$-chain is
a sum of the cost of creating the kink, $\ln(1/q)$, and of interaction with
magnetic field, $\sigma g(z)$, leading to the Boltzman factor
$q\exp(-\sigma g)$. Here the additive constant in the definition of the
potential $g$ is arbitrary but has to be the same for all kinks. One can see
now that (7) represents a single kink localized near a global extremum of $g$.
The uniform fixed points naturally correspond to the chain either fully
magnetized in one of the two possible directions, $\varepsilon=\pm 1$, or
maintaining zero average magnetization, $\varepsilon=0$, at $T_{d=1}=0$. The
low-$q$ corrections to these states represent a background of thermally
excited {\em single} kinks in (4), and {\em pairs} of kinks in (6). Other
fixed points do not have so simple interpretation, as kinks strongly interact
with each other there.

None of the fixed points considered so far addresses directly the situation in
which $t(z)$ has more than one zero, so that $g(z)$ has many maxima and minima.
A crucial insight is provided by consideration of the simplest system with {\em
two} boundaries. In the present formalism a strip of width $L$ is represented
by
 $t(z)=t_1$ at $|z|<L/2$, surrounded by $t=+\infty$ background
\cite{levfish2,onsager}. At $t>0$ the energy profile is asymptotically flat,
but at $t_1<0$ and for sufficient thick films, $|t_1|L\gg 1$, the contradiction
between the tendency to ordering within the film and the absence of any order
outside of it is resolved by creation of a pair of kinks separating ``cool''
interior from the ``hot'' background. Formally a solution is obtained by
matching two kink solutions (7) of opposite orientations with three noncritical
solutions (6) by choosing appropriate integration constants in those. This
construction collapses, however, when $q$ is decreased past
$\xi_{\parallel}^{-1}\propto e^{|t_1|L}$. In the kink picture this comes
naturally: the kink-antikink pair is annihilated as soon as the ``self-energy''
of the pair $2|\ln q|$ exceeds the potential difference
$|g(L/2)-g(-L/2)|=2|t_1|L$ supporting the pair. Physically \cite{onsager},
the exponential dependence of the crossover scale $\xi_{\parallel}$ on
$|t_1|L$ is a signature of the one-dimensional nature of the large scale
fluctuations in the film (recall $\ln\xi\propto T^{-1}$ in a classical Ising
chain) \cite{FP}: long range order in a strip of finite width is lost due to
activation of two-dimensional domain walls running across the film; the
Boltzman factor corresponding to creation of a pair of these walls is
$\exp[-2\Sigma(T)L/k_bT]$, where the free energy per unit length of the wall,
$\Sigma$, happens to be given by $|t_1|$ in this formalism \cite{onsager}.

I conjecture now that this mechanism of creation and annihilation of pairs of
kinks at maxima and minima of $g(z)$ defines the asymptotic RG flow for a {\em
generic} temperature field $t(z)$: at a given small $q$ a solution can be
constructed by matching pieces of (6) of different signs with the
kink-solutions (7). A pair of kinks is stable while these kinks are separated
by a potential difference $\Delta g>|\ln q|$; the pair is annihilated at
smaller $q$'s. This scenario correctly describes intermediate-scale behavior
in the exactly solvable periodic LPI superlattice \cite{onsager}. Because the
fluctuations of potential $g$ in the periodic model are strictly bounded, the
critical flow is ultimately attracted by the uniform critical fixed point (4).
On the other hand, a random, short range correlated $t(z)$ gives rise to
unbounded fluctuations of the random walk type:
$\overline{\Delta g^2(z)}=D|z|+o(z)$, with overbar standing for disorder
average. Consequently, at a given $q$ the typical size of a pair is going to
be $\Delta z=D^{-1}|\ln q|^2$, implying anisotropic scaling
$\xi_{\parallel}\propto\exp(D^{1/2}\xi_{\perp}^{1/2})$ at the {\em critical
random} fixed point; the subscripts $\parallel, \perp$ indicate directions
parallel and perpendicular to the layers. Consider now a random multilayer
grown by susequent deposition of two different magnetic materials: $t(z)$
takes two values, $t_1$ and $t_2,\  t_1<t_2$ with probabilities $p$ and $1-p$,
respectively. The average $\bar{t}=pt_1+(1-p)t_2$ goes through zero at a
certain temperature $T_c$; elsewhere a linear bias $\bar{t}z$ has to be added
on top of the random walk exhibited by by $g(z)$ at $T_c$. Crossover from the
critical random to one of the noncritical fixed points happens when potential
difference $\Delta g(\xi_{\perp})=\bar{t}\xi_{\perp}$ coming from the linear
part of $g$ exceeds the typical random fluctuation on that scale,
$(D\xi_{\perp})^{1/2}\sim\ln \xi_{\parallel}$, so that on larger scales $g$ is
mostly monotonic. This consideration gives
\begin{equation}
\xi_{\perp}=D\bar{t}^{-2},\ \xi_{\parallel}=\exp(D\bar{t}^{-1}),
\end{equation}
in agreement with the identification of the {\em average} correlation lengths
in [1(a)]; these lengths are different from the {\em typical} correlation
lengths, as disorder-average of the {\em logarithm} of the correlation
function results in $\tilde{\xi}_{\perp}\propto\bar{t}^{-1}$ \cite{SM,onsager}.
Although, for instance, at $\bar{t}>0$ the overall positive slope of $g$
dominates at scales larger than $\xi$, sufficiently long sequences of
$n\geq |\ln(q)|/|t_1|$ layers of pure low-temperature component occur with
small but nonzero density $p^n$. The associated rare fluctuations in the slope
of the potential keep a dilute system of large scale kink pairs from
annihilation at any $q$. First appearance of these large scale pairs at the
critical temperature $T_{c1}$ of the first component, where $t_1$ becomes
negative, leads to a Griffiths singularity in the free energy, which is easily
estimated in agreement with \cite{SM}; naturally, the same argument yields a
singularity at $T_{c2}$. One should note that while the singularity at $T_c$
depends only on the universal large-scale variation, $\Delta g\propto z^{1/2}$,
the essential singularities are governed by the tails of the distribution; if
the distribution of $t(z)$ is Gaussian the singularities are pushed all the way
to $T=0,\ \infty$.

This picture is easily generalized to the cases of long-range correlated random
and quasiperiodic LPI models. In the first case, the main difference comes with
the anomalous scaling dimension $x_g>\frac{1}{2}$ describing the fluctuations
$\Delta g(z)\propto z^{x_g}$ at criticality, $\bar{t}=0$. The correlation
lengths then diverge as $\xi_{\perp}\propto t^{-\nu_{\perp}},\
\xi_{\parallel}\propto\exp[\mbox{const}\ t^{-\tilde{\nu}_{\parallel}}]$, with
$\nu_{\perp}=1/(1-x_b),\ \tilde{\nu}_{\parallel}=x_b/(1-x_b)$. On the other
hand, if a quasiperiodic one-dimensional binary lattice is constructed by
projecting from a strip of finite width cut out of a two-dimensional lattice
\cite{LN}, then $g(z)$ can be obtained by a similar projection and turns out
to be confined within a strip of finite width around the average
$g\propto\bar{t}z$. Thus critical behavior in this ``Fibonacci'' multilayer is
governed by the pure Ising fixed point (4). Note that the amplitudes of the
long wave length harmonics $t_k\exp(ikz)$ of $t(z)$ in the Fibonacci sequence
are bounded by $t_k=O(|k|)$; in order for a quasiperiodic sequence to
extrapolate between periodic and random behavior, like it was found in
\cite{LN} for a quantum quasiperiodic XY-chain, the amplitudes have to go to
zero slower than $|k|$ so as to generate unbounded fluctuations in the
potential $g$.

In conclusion, an exact functional real space renormalization group flow has
been set up and studied for general layered planar Ising models. The RG scheme
allows to separate relevant and irrelevant features of a wide class of systems,
leading to the classification of various types of critical behavior based on
simple basic characteristics, such as the asymptotic variation of the thermal
potential $g=2\int tdz$. The surface and random geometries have been analyzed
within a single framework leading to a new physical picture of the criticality
in a general disordered LPI model: At any finite momentum transfer $q$ along
the layers the system behaves as an alternating sequence of domains of the
high- and the low-temperature phases. Having only one infinite dimension, the
domains undergo an infinite coarsening as $q$ decreases. The mechanism of
coarsening is activation of linear defects across the finite widths of the
domains, hence the exponential scaling (11) at the random criticality. The
large fluctuations in the bond strengths underlying the domain structure in
this formalism, must be essentially similar to those playing central role in
the RG approach of [1(a)].

Discussions with Daniel Fisher, Mark Oxborrow, and Heiko Rieger have been very
helpful.

\end{document}